# What Affects Team Behavior?
# Preliminary Linguistic Analysis of Communications in the Jazz Repository

Sherlock A. Licorish and Stephen G. MacDonell

*SERL, School of Computing & Mathematical Sciences*
*Auckland University of Technology*
*Private Bag 92006, Auckland 1142, New Zealand*
sherlock.licorish@aut.ac.nz, stephen.macdonell@aut.ac.nz

**Abstract**

*There is a growing belief that understanding and addressing the human processes employed during software development is likely to provide substantially more value to industry than yet more recommendations for the implementation of various methods and tools. To this end, considerable research effort has been dedicated to studying human issues as represented in software artifacts, due to its relatively unobtrusive nature. We have followed this line of research and have conducted a preliminary study of team behaviors using data mining techniques and linguistic analysis. Our data source, the IBM Rational Jazz repository, was mined and data from three different project areas were extracted. Communications in these projects were then analyzed using the LIWC linguistic analysis tool. We found that although there are some variations in language use among teams working on project areas dedicated to different software outcomes, project type and the mix of (and number of) individuals involved did not affect team behaviors as evident in their communications. These assessments are initial conjectures, however; we plan further exploratory analysis to validate these results. We explain these findings and discuss their implications for software engineering practice.*

## 1. INTRODUCTION

Since the emergence of software development as a discipline it has been plagued with contradictions over the adoption of specific procedures and tools, as well as inconsistencies in terms of projects' success rates [1, 2]. These outcomes result in speculation over which approaches and tools are more suitable, when, and for whom, and how they should be used to provide maximum value for the software development community [3, 4]. Despite ongoing efforts to improve software development practices through such initiatives, uncertainties over project success rates remain [5]. This has led in part to a growing belief that software development performance would improve more substantively if the human processes employed during this activity were better understood and supported [6-8].

One avenue for understanding these processes is to study the artifacts produced by those working on software projects. It has been shown previously that software artifacts and software history data are useful sources of interaction evidence (e.g. see [9]). Communication artifacts such as electronic messages, change request histories and blogs have provided unique perspectives on activities occurring during the software development process [10, 11]. While the study reported here is a retrospective one, in-process interrogation of such artifacts could be useful in proactively managing interactions during software development projects. With this intent in mind we used data mining and linguistic analysis in a preliminary study of team behaviors using the IBM Rational Jazz repository. Our research agenda, approach and study findings are reported here.

In the next section we consider prior related work, establish the basis for our research approach and outline our specific research question. We then describe our research methods and study context in Section 3, and we also introduce our linguistic analysis measures in this section. In Section 4 we present our results and initial analysis. Section 5 comprises a discussion of our findings and outlines implications of our results as well as threats to validity, and in Section 6 we draw our study conclusions.

## 2. BACKGROUND AND RELATED WORK

Understanding the human processes and team behaviors that occur in software development may be facilitated by studying evidence of communication and coordination. Coordination involves connecting and managing resources – primarily personnel – and the interdependencies between activities and tasks [12, 13]. In software development, coordinating activities among team members possessing diverse backgrounds (including language, culture and sometimes over geographical distance) is inherently challenging [14] and if not keenly managed during problem-solving tasks may lead to longer development times [15] and larger numbers of defects [16].

Thus, careful synchronization between and among those involved in software development becomes paramount in task accomplishment. It is crucial that changes in tasks and their dependencies are reported to all involved in a timely manner to avoid task failures [17]. While there are many methods for coordinating and conveying changes, including official documentation and communication between participants [12], as well as tools to support change coordination [18], previous work has shown that documentation generally becomes obsolete as a software project progresses and, most often, the team members themselves are the main sources of current knowledge (shared through their communication) [19, 20]. In fact, an early study (1994) exploring software developers' activities found that up to 50 percent of practitioners' time was spent on interpersonal communication and coordination during software problem solving [21]. Studying these interactions therefore has the potential to reveal the reasons for, and consequences of, communication, coordination and action during software development projects.

This view is supported by the extent of prior research that has been dedicated to this subject, software teams' communication and coordination having received a growing amount of attention in the literature [22-24]. A case study approach adopted by Damian and colleagues [25] at a large IT manufacturing company in Brazil unearthed that software programmers interacted more than any other group considered in the study. In an earlier investigation of collaboration around system requirements, Potts and Catledge [26] found that final agreement and acceptance of a requirement necessitated participants' constant reorientation, that knowledge was often lost between stages, and that conflict existed between those implementing the project and those managing the project. A study of coordination conducted by Ehrlich et al. [27] found that brokers bridge communication gaps for teams that communicated across distributed sites. Using online surveys, Chang and Ehrlich [28] identified that team leaders often acted as coordinators in software development and team communication positively influenced team awareness and team climate.

While these studies have provided insights into the reasons for and functions of team communication, they did not provide cues from an internal project perspective. Various behaviors and traits may be necessary and prevalent in some software environments or contexts, while other settings may demand different attitudes for teams to succeed. The absence of these specific arrangements may throw out team balance and result in challenges to the success of the software project. Such a position is supported by work on role theories. Role theories have identified both positive and negative group behaviors in teams, and indicate that these must be balanced if teams are to succeed. In their seminal study of roles in groups, Benne and Sheats [29] found that team social interaction is one of the key influential factors of success in group work. These authors observed evidence of team roles that promote helpful and supportive behaviors (personal and social roles), task-concerned behaviors (task roles), and debate- and conflict-centered behaviors (individualistic roles).

Belbin [30] found that in successful teams nine roles are performed by team members. Similarly to Benne and Sheats [29], Belbin contends that a person's interactions in a group are influenced by their natural behavioral preference(s). Among his findings he also reported that individuals who possessed exceptional quality in one respect (e.g. social) may demonstrate weakness in other traits (e.g. idea generation), and that bringing together individuals with similar preferences is likely to lead to reduced team performance. Belbin established that successful teams are heterogeneous; normally possessing a balance of team members occupying all roles (noting that individuals can possess preference for more than one role, having a primary preference and other secondary preferences). Individuals are most comfortable when they are functioning in roles that are their natural preference. Interaction between different roles, without understanding them and managing their differences, can be a source of team conflict.

Outside of the role theories domain, work in human resource management has also integrated psychology and role theories in supporting the task of selecting individuals with appropriate skill sets for positions. In particular, most software-related positions demand multiple capabilities, including intra-personal, organizational, inter-personal and management skills [31, 32]. Intra-personal skills include judgment, innovation and creativity, and tenacity, while being self-organizing and having knowledge of specific environments (e.g. programming competences in Java or Microsoft technologies - which may be supported by training) is characterized as organizational. Inter-personal skills comprise team work, and cooperation and negotiating skills, and management skills are related to planning, organization and leadership.

According to linguistic theories, it is possible to discern these skills within individuals' communications [33]. Linguistic studies have shown that individual language use is stable over periods of time and the way individuals communicate is also influenced by their context and local settings [34].

Language use has also been studied as a function of age [35], gender [36] and emotional upheavals [37]. These studies provide compelling evidence that language use is contextual. Thus, studying team members' communications from a linguistic perspective during software projects may help us to understand - does the project environment (project type, people involved) affect team behaviors as evident in their communications?

## 3. METHOD AND MEASURES

We employed a multiple case study design in this preliminary analysis of the IBM Rational Jazz Repository. Jazz is a fully functional environment for developing software and managing the entire software development process, incorporating project management, project communication and development [38]. The environment includes features for work planning and traceability, software builds, code analysis, bug tracking and version control in one system [39]. Changes to source code in the Jazz environment are allowed only as a consequence of earlier tasks created, such as a bug report, a new feature request or an existing feature amendment. Features and artifacts are tracked using work items (WIs), and a WI represents a single task which may be a defect repair, an enhancement or a development task. Team member communication and interaction around WIs are facilitated by Jazz's comment functionality. We were given access to a large amount of software development data from activities undertaken by teams spread across the United States, Canada and Europe. This instance of the repository (release 1.0.1) includes numerous projects (see products at jazz.net), with specific teams responsible for various project outcomes. It is also not uncommon for team members to work across many teams.

We worked deliberately to ensure that interesting variations in the repository were captured in the data sampling. Since the aim of the study was to assess project environment and team behaviors from communication data, the most desirable information-rich cases (project areas) were those in which there was a high intensity of messages around features to be delivered. Additionally, our goal was to select cases that represented the scope and breadth of the various project areas in the repository, for example: addressing user experience, coding and project management-oriented activities.

A Java program was created to leverage the Jazz Client API to extract the required development and communication artifacts. These included Work Items representing project management and development tasks, Project Workspaces representing multiple project areas and Comments representing practitioner dialogues. Extracted project information included 36,672 resolved work items (from the various project areas in the repository) created between June, 2005 and June, 2008. We extracted 94 project categories that comprised more than 25 work items, providing potential support for the planned investigations. The team workspaces consisted of 474 active contributors belonging to eight different roles. Practitioner comments – our primary data source – were also extracted, totaling 117,571. The data extracted from Jazz were imported into a relational database management system to facilitate efficient data management.

As stated above, of this wealth of information we mined the data from three project areas, characterized by the summary measures shown in Table I. The Linguistic Inquiry Word Count (LIWC) tool was then used to analyze practitioners' communications with the intent to provide insights into team behaviors in terms of the perspectives listed in Table II [40].

The LIWC is a software tool created after four decades of research using data collected across the USA, Canada and New Zealand [41, 42]. Data sources used in creating the LIWC tool spanned many areas of life, including emotional writing, control writing, research articles, blogs, novels and normal conversations (and data collection is an ongoing exercise). This tool captures over 86% of the words used during conversations (around 4500 words) and is available in many languages. In this tool words are grouped into specific categories, such as negative emotion, social words, and so on (see Table II). Written text is submitted as input to the tool in a file and is then processed and summarized based on the mapping of input source words to those in the LIWC dictionary. As each word searched for in the LIWC dictionary is found, the associated scales are incremented based on the word category, after which a file is returned with the summary output. The output data include the percentage of words captured by the dictionary, standard linguistic dimensions (which include pronouns and auxiliary verbs), psychological categories (cognitive, social) and personal dimensions (work, achieve, leisure and so on). Table II describes the LIWC linguistic measures that were analyzed during our exploration, along with brief justifications for their inclusion based on their relevance to our research goals.

**TABLE I**. Summary Statistics for the Chosen Jazz Project Areas

(P1 tasks related to developing UIs, P2 tasks were under the project managers' control, P3 tasks were associated with middleware development)

| Project Area ID | WI /Task Count | Project Category | Team Size | Total Messages | Period (days) |
|---|---|---|---|---|---|
| P1 | 54 | User Experience | 33 | 460 | 304 |
| P2 | 210 | Project Management | 90 | 612 | 660 |
| P3 | 207 | Code (Functionality) | 48 | 640 | 520 |

**TABLE II**. LIWC Linguistic Measures Used in This Study

| Linguistic Category | Abbreviation | Examples | Reason for Inclusion |
|---|---|---|---|
| Pronouns | i | i, me, mine, my | Individuals favoring more collective group process may demonstrate this trait through their language use [43]. Previous research has found elevated use of first person plural pronouns (we) during shared situations and among individuals that share close relationships, whereas, relatively excessive use of self references (i) has been linked to individualistic attitudes [33, 44]. First person singular and plural pronoun linguistic dimensions are considered here to analyze shared group processes among members. Use of the second person pronoun (you) may signal the degree to which members rely on (or delegate) other team members or their general awareness [33] of others and their activities. This phenomenon is assessed by assessing the second person plural pronoun linguistic dimension. |
| | we | we, us, our, we've | |
| | you | you, your, you'll, you've | |
| Cognitive language | insight | think, believe, consider, determined | Software teams were previously found to be most successful when many group members were highly cognitive and natural solution providers [45]. These traits also previously correlated with effective task analysis and brainstorming capabilities. These linguistic dimensions are included so that we can analyze communication artifacts to assess the cognitive aspects of team members. |
| | discrep | should, would, could, prefer | |
| | tentat | maybe, perhaps, apparently, chance | |
| | certain | definitely, commit, always, extremely | |
| Work and Achievement related language | work | feedback, goal, boss, inventory | Individuals most concerned with task completion and achievement are said to reflect these traits during their communication. These individuals are most concerned with task success, contributing and initiating ideas and knowledge towards task completion [29]. Work and achievement related communication are analyzed to assess those most concerned with task completion. |
| | achieve | accomplished, resolved, obtained, finalized | |
| Leisure, social and positive language | leisure | movie, artist, party, play | Assessment of the use of leisure terms, the opposite to work, is used to measure the relative frequency of off-task interactions within teams. Individuals that are personal and social in nature are said to communicate positive emotion and social words and this trait is said to contribute towards an optimistic group environment, promoting encouraging, harmonizing and compromising traits [29, 46]. |
| | social | gossip, give, buddy, love | |
| | posemo | beautiful, relax, perfect, glad | |
| Negative language | negemo | afraid, bitch, hate, suck | Negative emotion may affect team cohesiveness and positive group environment. Those expressing significant negative emotion are also said to have a tendency to show excessive anger [47]. |

## 4. RESULTS AND INITIAL ANALYSIS

Figure 1-A shows that all three project areas exhibited relatively low levels of individualistic language, but even lower levels of collective language. Use of reliance and delegation terms was also low for all teams, with those communicating on user experience projects (P1) utilizing slightly more of this language form. Figure 1-B shows that evidence of the cognitive dimensions was low overall, but those communicating on the project management-oriented tasks (P2) were both more insightful and more tentative than those working on the other projects. There was also very low certainty exhibited in communications for all teams (see Fig. 1-B). Figure 1-C shows that teams in all three areas were concerned about task completion and achievement, and that teams spent very little time communicating about leisure. Those working on the user experience projects (P1) communicated with positive language, while those working on the project management projects (P2) were the most social in terms of their communications (see Fig. 1-D). Overall, positive and social language use levels were much higher than that of negative language use, Fig. 1-D also showing that all teams communicated less than five percent negative emotion overall (with slightly higher use of this language form on the middleware development activities).

We divided each project area equally into four time periods to assess variations in team behaviors over project duration. Observations (see Fig. 2) for the three project areas are that individualistic language use tended to be most evident in the earlier phases projects and decreased as projects progressed (apart from in the last P2 phase). Social language use was highest at the start of projects and then tended to fluctuate throughout the remaining phases. Software practitioners' communications were also most insightful at the start of projects. In contrast, the work and achievement language dimension fluctuated but was most evident during the later phases of the software projects (the second phase of P1 aside).

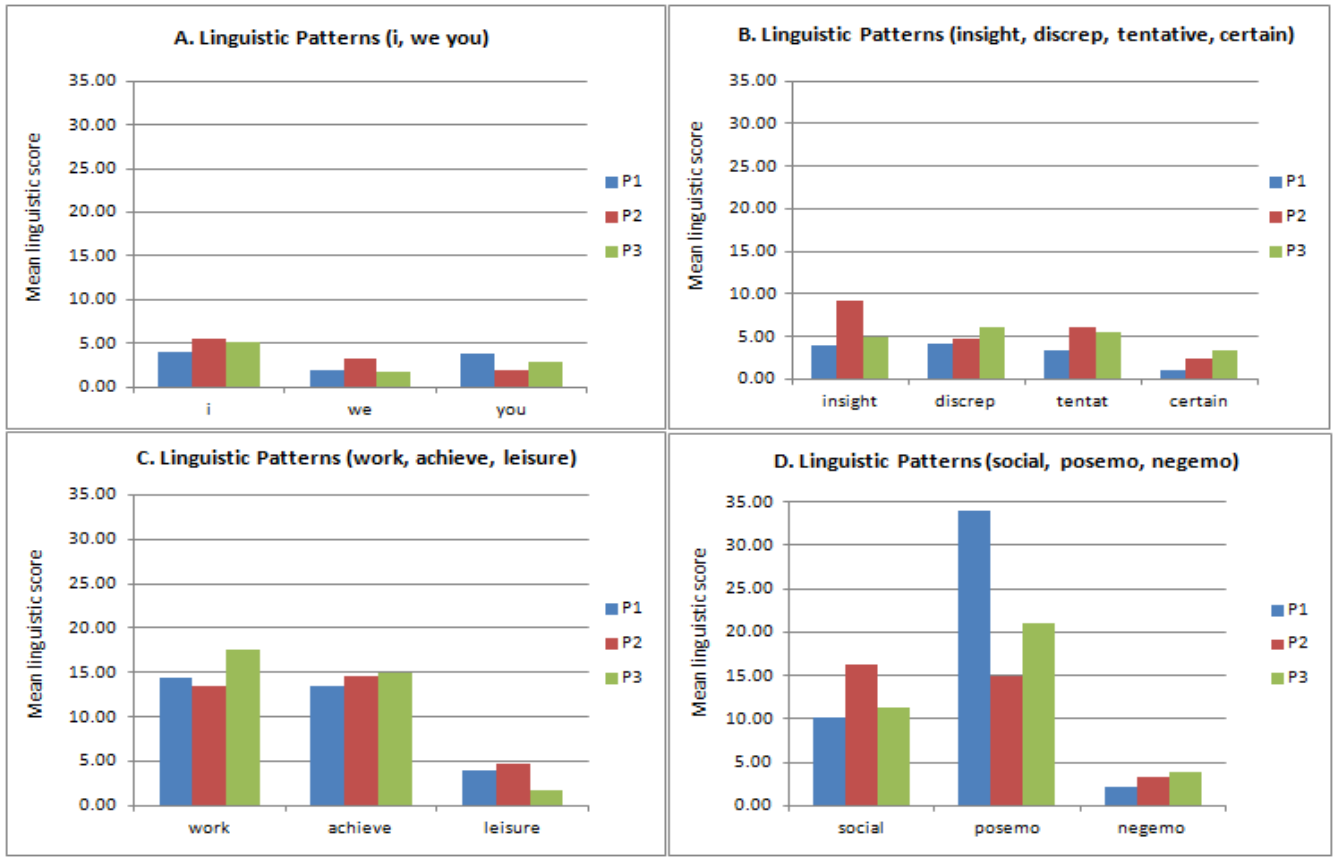

**Figure 1**. Mean linguistic scores (percentage use) for User Experience (P1), Project Management (P2) and Coding (P3) projects

## 5. DISCUSSION

Does the project environment (project type, people involved) affect team behaviors as evident in their communications? Our analysis suggests that, regardless of the project type or people involved, Jazz team members used minimal collective, individualistic and delegation language. In fact, there was slightly more individualistic language use than collective and delegation language. These findings are of interest given the highly collaborative and experienced nature of the distributed Jazz teams [48], as we had assumed that such teams would utilize very collective group processes. The slightly higher levels of individualistic language may reflect that specific tasks (for instance, defect resolution) are assigned individually rather than collectively, and a more collective approach may be evident among those working on new software features. Contrary to some previously held views [49], the measures observed for the cognitive dimensions in these projects were not especially prominent. Software developers are generally held to be strong thinkers and creative individuals. Thus, there was an expectation that these practitioners would exhibit relatively high levels of use of cognitive processes during their discourses (given the IBM Rational product base available at jazz.net), but this was not evident. This may signify that such expertise may not be easily revealed in communication or the LIWC tool did not entirely capture the specialized terminologies used in this software development context.

The project management teams' mean score for insightful language was double that for those working on the user experience and coding projects, suggesting that this team encompassed more insightful individuals and/or that the project lent itself to encouraging such comments. Members of all teams were concerned with task achievement but were less vocal regarding leisure. Compared to those involved in the other two projects areas, team members working on the user experience projects used a very high proportion of positive language. While this may be linked to task type – those designing user interfaces may use friendly language due to the nature of their job e.g., "this interface looks nice", "I like these colors" – these observations may also indicate a requirement for more social or people-oriented behaviors among those working on user experience related tasks. Positive and social traits are said to contribute positively to group environments, while the opposite is said about negative emotion, which was low for all teams.

Overall, our identification of similar values for most of the linguistic dimensions across the three project areas may support conjectures that project type does not influence behaviors as evident in the language used in the Jazz developers' communications (leading to us proposing the model in Fig. 3) or that the Jazz teams are homogenous. In fact, we considered the similarities in the projects' team membership and team members' overall contributions to their project dialogue and found that although 16 members (of more than 170) were common in all three project areas, their engagement varied on the projects (e.g., those that contributed 67% of the project communication in P1 contributed only 1% of the project messages in P2, and 4% of the exchanges in P3, and this trend was maintained for the main contributors on other projects). These dissimilarities among the common members' participation in the three project areas provide some support for our proposition. Additionally, higher measures observed for work, achievement, social and positive linguistic processes support Benne and Sheats' [29] theory on the necessity for both social and task roles during team work.

The finding that individualistic language use decreased as projects progressed could be interpreted to mean that team processes became more collective and matured over time (at least in this case considering the Jazz team members). Additionally, it is perhaps understandable that social language use would be at its highest at the start of a project, as team relationships were formed, and that work- and achievement-related language use would be highest towards project completion, given the expected pressures on teams at delivery time.

## A. Threats to Validity

We acknowledge that there are shortcomings to this study that may pose threats to the work's validity.

*Construct Validity*: Construct validity reflects the adequacy with which variables represent the intended construct of interest [50]. The linguistic constructs used here to assess team behaviors have been used previously and were tested in prior studies for validity and reliability [33, 43]. However, the adequacy of these constructs in this particular study context has not been evaluated. Additionally, communication was assessed only from messages sent regarding software tasks. These messages were extracted from Jazz itself, and may not represent all of the projects' communications, some of which may have occurred through email and chat, through other tools used in the environment, as well as via face-to-face for collocated team members. Nonetheless, previous work has confirmed that the method of communication studied here is widely used for project discourses during software development at IBM [48] and so is valid in that context.

*External Validity*: The work processes at IBM Rational are specific to that organization and may not reflect the organization dynamics in other software development establishments. The software teams studied in this work used Jazz for project execution, including project management, project communication and software building, and followed specific software methods. These software processes may differ from those used in other software organizations or in open source project environments. Thus, given the small sample size considered here, and that other software teams may not create software in a similar environment to that in IBM Rational, the results found through this work may not necessarily generalize to all Jazz teams or to other software development situations.

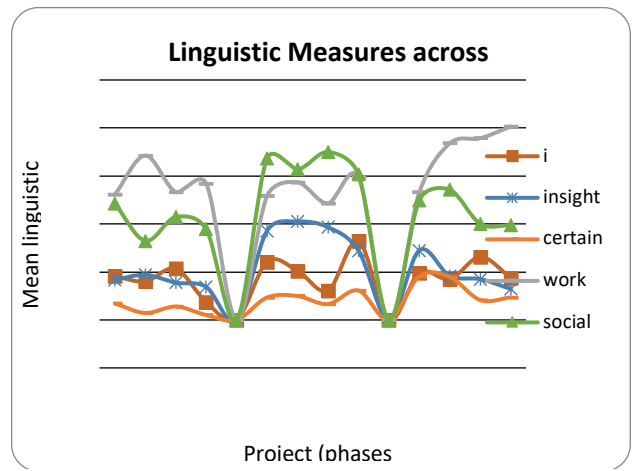

**Figure 2**. Linguistic measures across project phases for User Experience (P1), Project Management (P2) and Coding (P3) projects

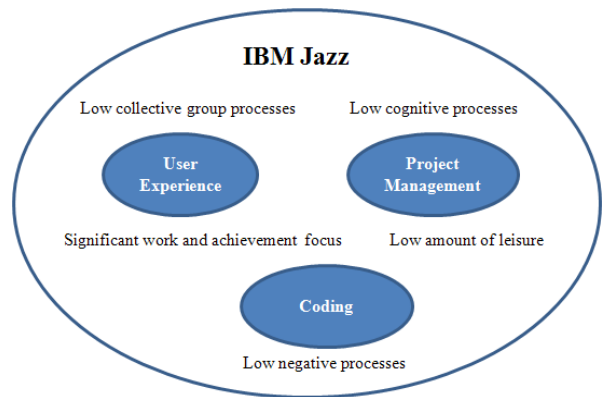

**Figure 3**. IBM Rational Jazz Team Behaviors model

# 6. CONCLUSIONS

In accepting the view that software repositories possess interaction evidence that may help us to understand the intricate human nature of software development, we conducted a preliminary study of team behaviors as evident in communication data for three project areas from the IBM Rational Jazz repository. Our results indicate that these Jazz teams communicated with relatively low levels of collective or cognitive language. Additionally, members of all teams were concerned about task achievement but not about leisure. Team members working on user experience projects used a higher proportion of positive language than others, and all teams used low levels of negative language. Overall, we observed similar findings for most of the linguistic dimensions across all three project areas, supporting the conjecture that the project type and the mix of (and number of) individuals involved did not affect team behaviors as evident in their communications. We also observed high levels of task-centric and social and positive behaviors, indicating that both traits are indeed necessary during group work. It would be useful to examine a larger sample of projects (and organizations) to see if these findings hold across a wider range of projects.

# ACKNOWLEDGMENTS

We thank IBM for granting us access to the Jazz repository. Thanks to Jacqui Finlay and Andy Connor for help setting up and mining the repository. S. Licorish is supported by an AUT VC Doctoral Scholarship Award. IBM and Jazz are trademarks of IBM Corporation.